\documentclass[12pt]{article}
\usepackage{amssymb}
\usepackage{epsfig}
\usepackage[
            body={14.6true cm,22true cm},
            twosideshift=0 pt,
            ]{geometry}
\begin{document}

\title{\bf A self-organized particle moving model on scale free network with $1/f^{2}$ behavior}

\author{{YunFeng Chang\thanks{Email: yunfeng.chang@gmail.com}, Long Guo, Xu Cai}\\
{\small Complexity Science Center,Institute of Particle Physics}\\
{\small Huazhong(Central China) Normal University, Wuhan, 430079,
China.}} \maketitle

\begin{abstract}
In this paper we propose a self-organized particle moving model on
scale free network with the algorithm of the shortest path and
preferential walk. The over-capacity property of the vertices in
this particle moving system on complex network is studied from the
holistic point of view. Simulation results show that the number of
over-capacity vertices forms punctuated equilibrium processes as
time elapsing, that the average number of over-capacity vertices
under each local punctuated equilibrium process has power law
relationship with the local punctuated equilibrium value. What's
more, the number of over-capacity vertices has the bell-shaped
temporal correlation and $1/f^{2}$ behavior. Finally, the average
lifetime $L(t)$ of particles accumulated before time $t$ is
analyzed to find the different roles of the shortest path
algorithm and the preferential walk algorithm in our model.
\end{abstract}

\bigskip

\section{Introduction}
Since the small-world network was proposed by Watts and Strogatz
in 1998\cite{s1} and the scale-free network was proposed by Albert
and Barab¨¢si in 1999\cite{s2}, much work on complex network
emerges, ranging from analyzing the topology of many real complex
systems and finding some universal characteristics of
them\cite{s3,s4,s5}, to modelling dynamics about the complex
systems and dynamical processes on complex
networks\cite{s6,s7,s8}. The study of complex network covers many
fields, such as sociology, chemistry, biology, physics and
computer science.

With the development of science and technology, dynamical
communication and information exchange processes on networks, such
as the Internet, the World Wide Web (WWW) and the traffic network,
are becoming more and more important. Many researchers have
studied the congestion phenomenon of information packets (data
packets and messages) on complex network, such as communication on
network with hierarchical branching\cite{s9}, the crossover
behavior and congestion/decongestion on a two-dimensional
communication network with regular vertices and
hubs\cite{s10,s11}, dynamics of jamming transition\cite{s12} and
phase transition on computer network\cite{s13}. On the one hand,
many authors studied the relationship between the dynamical
systems and their underlying network structure\cite{s14,s15} to
find some new routing strategies to improve the transportation
efficiency on complex networks \cite{s16,s17}. Much attention is
paid to the  study of the global statistical properties of dense
traffic of particles on scale free network\cite{s18}. On the other
hand, there also appears some work\cite{s19,s20} which focus on
the individual transportation behavior of each vertex of the
network since the time dependent activity of each vertex captures
the network transportation system's dynamics from a different
angle, and those parallel time series can increasingly complete
the information about the system's collective behavior.

In this paper we propose a self-organized particle moving model on
scale free network with the algorithm of the shortest path and
preferential walk. We study the over-capacity property of the
vertices in particle transportation system on complex network and
find that the number of over-capacity vertices $A(t)$ has the
punctuated equilibrium property, bell-shaped temporal correlation
and $1/f^{2}$ behavior. We also investigate the average lifetime
of particles accumulated before time $t$ under different
parameters, which can indicate directly the different roles of the
shortest path algorithm and the preferential walk algorithm in our
model.

This paper is organized as follows. In section 2 we give out our
model in details, in section 3 the simulation results and finally
in section 4 our conclusion.

\section{The Model}

The network here is described as undirected and unweighted graph
$G=(V,E)$\cite{s15}, where $V$ is the set of vertices, and $E$ is
the set of edges among the vertices. Multiple edges between the
same pair of vertices, as well as loop which is an edge beginning
from and ending at the same vertex are not allowed. We generate
the scale free network by using Barab$\acute{a}$si and Albert's
algorithm of growth and preferential attachment\cite{s2}, but
start from a random network with $m_{0}$ vertices and $l$ edges.
Then, vertex with $m(m<m_{0})$ edges is attached iteratively.
Fig.1 is the degree distribution of the underlying network of our
particle moving model.

\begin{figure}[th]
\centering
\includegraphics[width=6cm]{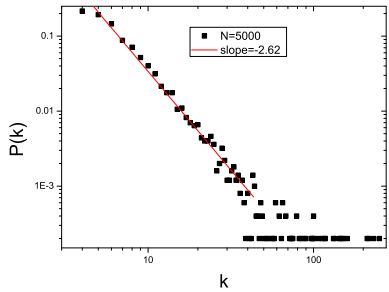} \vspace*{0pt}
\caption{the degree distribution of the underlying network of our
particle moving model}
\end{figure}

In the previous works, there are two kinds of different
definitions for load. One is that load is independent of the
dynamical process and is the same as the betweenness of the vertex
that defined in \cite{s21} as the number of the shortest paths
connected arbitrary pair vertices in the network that pass through
the vertex. The other is dependent on the searching process and is
defined as the number of particles that pass through the vertex or
the number of particles that stay at the vertex presently. In this
paper, we utilize the later definition that the load of a vertex
is the number of particles staying at the vertex presently, which
varies as time elapsing. We define one physical quantity: capacity
$C_{i}$ as the maximal load of vertex, which is proportional to
its degree with a tunable parameter $a$, so that the capacity of
the network has the same distribution as the degree of the network
which reflects the topology of the underlying network.

Each vertex has, for one thing, the ability of generating and
delivering particles which will move on the network, and for
another, two states which are normal and over-capacity
respectively. If the load of a vertex is larger than its capacity,
we call this vertex over-capacity. Every time step, $n$ particles
are generated. The source of each particle and its destination are
chosen randomly among all the vertices of the network. Besides,
for simplicity and convenience, each vertex sends only one
particle each time. The vertex $i$ of the network can have a queue
of $L_{i}(t)$ particles that are waiting to be delivered, which is
the load of the vertex $i$ at time $t$ as defined above.

Particles moving on the scale free network obey the shortest path
algorithm\cite{s11,s12,s14} and the preferential walk algorithm.
Take a particle $P(i,j)$ which is going from vertex $i$ to vertex
$j$ at time $t$ for example, if the next vertex $k$ from vertex
$i$ to vertex $j$ along the shortest path is normal
$L_{k}(t)<C_{k}$, the particle will wait at the end of the queue
at vertex $k$ until the particles ahead leave from vertex $k$.
While if vertex is over-capacity $L_{k}(t)$$\geq$$C_{k}$, the
particle will utilize the preferential walk algorithm to go to the
front of the queue at vertex $k$, and move to one of vertex $k$'s
nearest neighbors which have the minimum product of degree and
load until there is no over-capacity vertices in the network.

\section{Simulation Results}

We realize our model on a scale free network with $N=500$
vertices, power law exponent $\gamma=-2.45$, clustering
coefficient $C=0.0435$, characteristic path length $L=3.18$ and
the average degree $<k>=8$.

\begin{figure}[th]
\centering
\includegraphics[width=6cm]{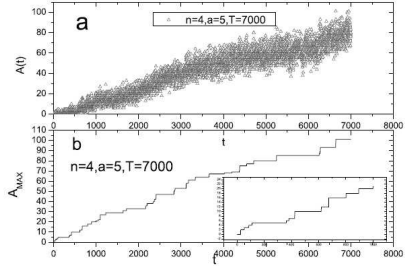} \vspace*{0pt}
\caption{the number $A(T)$ of the over-capacity vertices vs. the
time t(a) and the punctuation equilibrium formation of $A(T)$(b)
with the parameter $n=4,a=5,T=7000$(the inside picture in picture
b is the first 1000 time steps)}
\end{figure}

One of the most interesting properties of our dynamical traffic
model is that the states of vertices  vary as time $t$ changes.
Hence, we study the relationship between the number of
over-capacity vertices $A(t)$ and time $t$ (see Fig.2). We can see
that, as time running and the number of particle increasing and
moving, the number of over-capacity vertices $A(t)$ has the
punctuated equilibrium behavior (see Fig.2 (b)). That means the
evolution of $A(t)$ can be divided into different time intervals.
After one time interval's accumulation, $A(t)$ will jump to a
higher value just like the update of life, e.g. in the former time
interval $\Delta t(=t_{2}-t_{1})$, there is no other
$A(t)(t_{1}<t<t_{2})$ that is bigger than $A(t_{1})$, then after
$\Delta t(=t_{2}-t_{1})$'s accumulation, at time $t_{2},
A(t_{2})>A(t_{1})$. We call this local punctuated equilibrium
process from $t_{1}$ to $t_{2}$ the local punctuated equilibrium
at $A_{max}=A(t_{1})$, and the next local punctuated equilibrium
is the local punctuated equilibrium at $A_{max}=A(t_{2})$.

We define the average number of the over-capacity vertices $\rho
(A_{max})$ under each different local punctuated equilibrium time
interval as the over-capacity density at $A_{max}$. Accordingly:

\begin{equation}
{\rho(A_{max})=\frac{\sum\limits_{t=t_{1}}^{t_{2}}A(t)}{\Delta t}}
\end{equation}
where, $\Delta t=t_{2}-t_{1}$, $A(t)<A_{max}=A(t_{1})$, $t_{1}\leq
t<t_{2}$.

\begin{figure}[th]
\centering
\includegraphics[width=6cm]{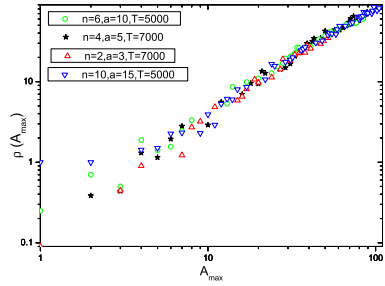} \vspace*{0pt}
\caption{the density of the over-capacity vertices $\rho(A_{max})$
vs. $A_{max}$ under different $n$ and $a$}
\end{figure}

We study the relationship between the over-capacity density
$\rho(A_{max})$ and $A_{max}$(see Fig.3). We can see that
$\rho(A_{max})$ has the same evolution trend as the function of
$A_{max}$ under different parameters $n$ and $a$. At the last part
of the graph, $\rho(A_{max})$ and $A_{max}$ reach the power law
relationship with the same slope, which indicates, in a
self-organized network transportation system, the crucial roles of
the interaction between the dynamical algorithm and its underlying
network structure.

What's more, we consider the number of over-capacity vertices as
the pulse signal which is the function of time. The so-called
signal can be characterized by the temporal correlation
function\cite{s22}:

\begin{equation}
G(t)=<A(t_{0})A(t_{0}+t)>_{t_{0}}-<A(t_{0})>_{t_{0}}^{2}
\end{equation}

$G(t)$ reflects the strength of statistical correlation between
the signal at time $t_{0}$ and the signal at time $t_{0}+t$, if
there is no statistical correlation, we have $G(t)=0$. On the
other hand, the temporal correlation function is related to the
power spectrum through a cosine transform as follows\cite{s22}:

\begin{equation}
S(f)=2\int\limits_{0}^{\infty}dtG(t)\cos(2\pi ft)
\end{equation}

\begin{figure}[th]
\centering
\includegraphics[width=5cm]{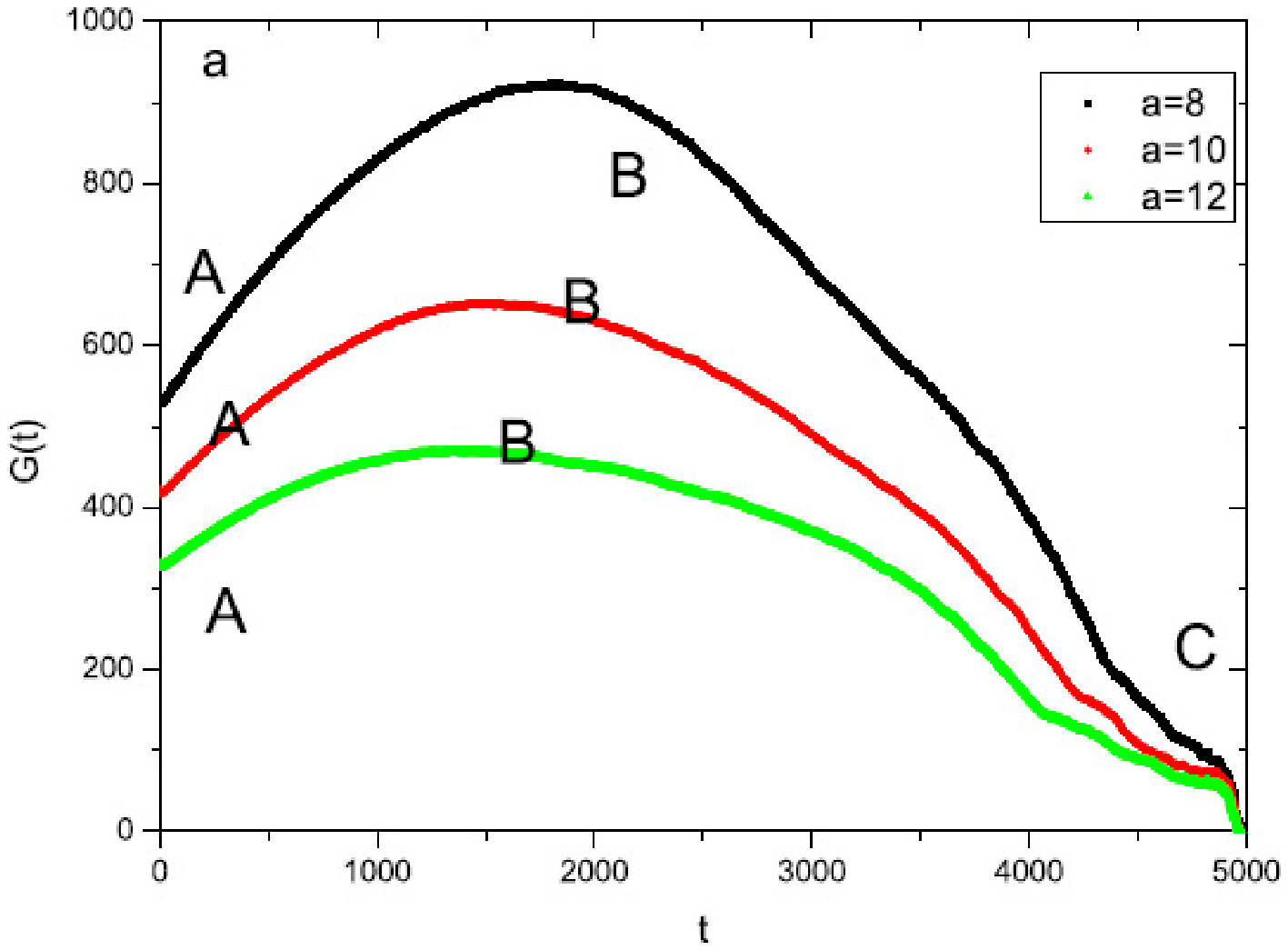} \vspace*{0pt}
\includegraphics[width=5cm]{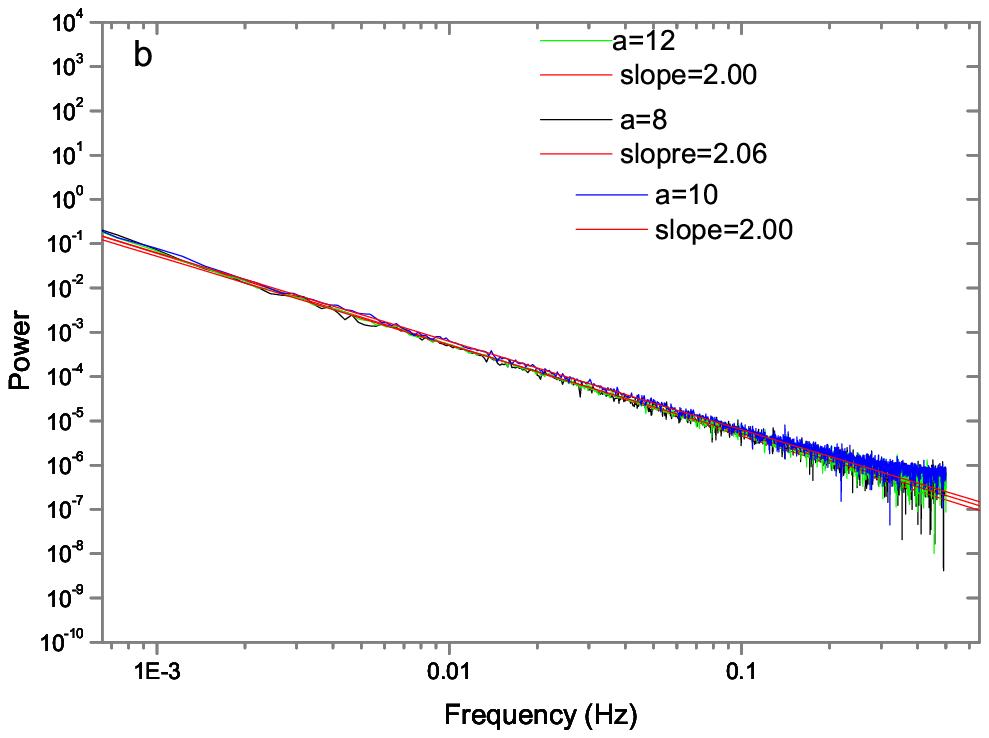} \vspace*{0pt}
\includegraphics[width=5cm]{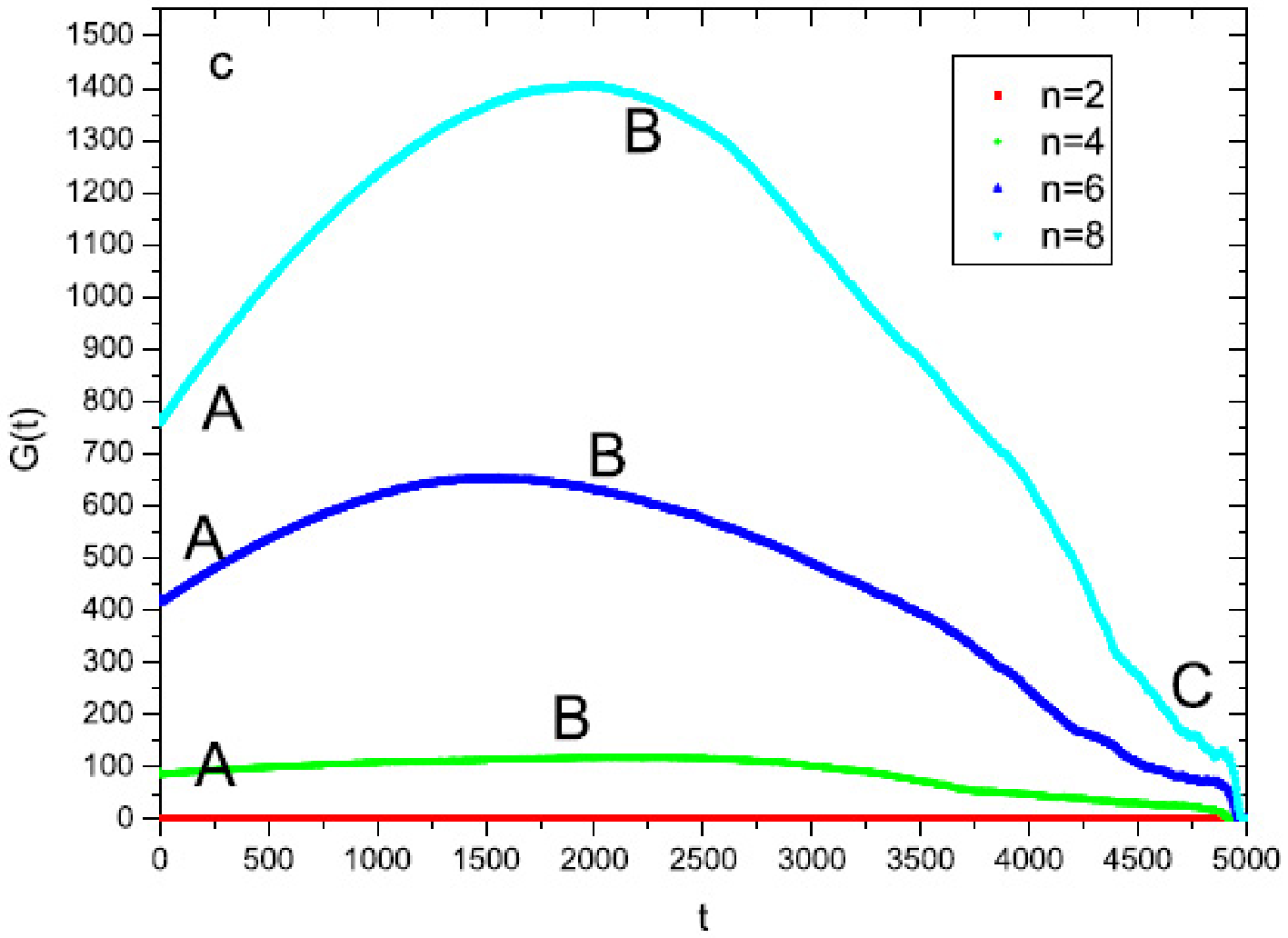} \vspace*{0pt}
\includegraphics[width=5cm]{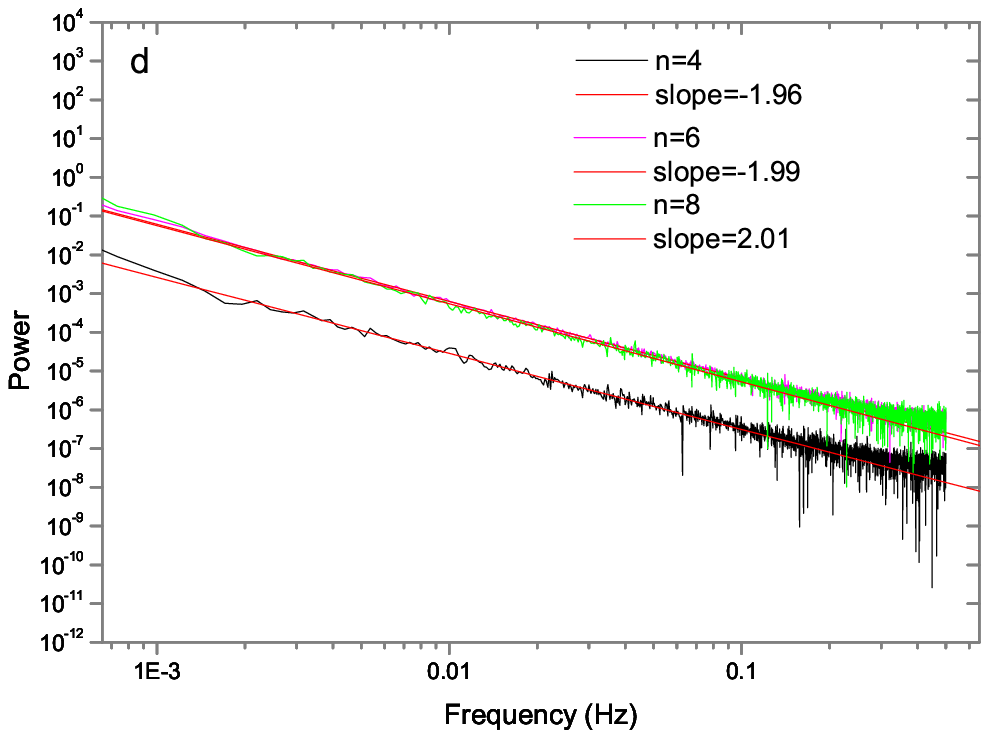} \vspace*{0pt}
\caption{temporal correlation of $A(t)$ and its power spectrum
accordingly}
\end{figure}

We study the behavior of the number of over-capacity vertices
$A(t)$ under different parameters by utilizing the temporal
correlation function and power spectrum (see Fig.4). The results
show that they all have bell-shaped temporal correlation and the
correlation strength is proportional to n and is inverse
proportional to a (see Fig.4a and Fig.4c). This again indicates
the crucial interaction, in a self-organized network
transportation system, between the dynamical algorithm and its
underlying network structure which is reflected very well in our
model.

On the other hand, the temporal correlation graph has the bell
shape which has symmetry property to some extent. We separate the
whole curve into two parts and label them AB and BC respectively
(see Fig.4a and Fig.4c). In part AB the shortest path algorithm is
the dominant dynamical algorithm and in part BC is the
preferential walk. We study the subsection power spectrum of the
temporal correlation. In this paper, we consider the BC section of
the temporal correlation under different parameters n and a, and
get the power spectrum accordingly (see Fig.4b and Fig.4d). We can
see that they have the same power law exponent that is independent
of the parameters n and a, and that the power $S(f)$ and the
frequency $f$ scale as:

\begin{equation}
S(f)\propto\frac{1}{f^{2}}
\end{equation}

In fact, the absolute value of the power spectrum of AB section on
log-log plot is the same as BC section, which indicates the
$1/f^{2}$ behavior is due to the interaction between dynamical
system and the underlying network structure but may have stronger
relationship with the topology structure of the underlying
network.

\begin{figure}[th]
\centering
\includegraphics[width=6cm]{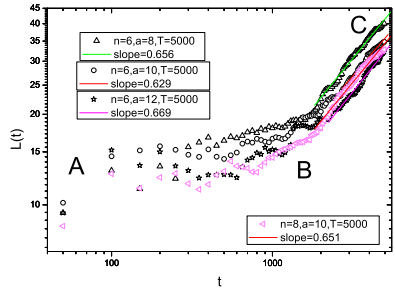} \vspace*{0pt}
\caption{the lifetime $L(t)$ of particles vs. $t$ under different
parameters $n$ and $a$}
\end{figure}

Finally, we analyze the average lifetime $L(t)$ of the particles
accumulated before time $t$, which can directly reflect the
different roles of the shortest path and preferential walk that
routes on the particle moving in our model(see Fig.5). In Fig.5,
the curve can be separated into two parts (AB and BC). The
lifetime of part AB varies slowly as time running in logarithmic
coordinate, which is caused by the particles that move on the
scale-free network according to the shortest path algorithm and
the statistical property. On the other hand, as the accumulation
of particles moving on the network, more and more particles move
according to the preferential walk, which causes the lifetime of
particles increases as time elapsing(BC part in the Fig.5). The
lifetime indicates that the dynamical algorithm plays an important
role in the dynamical phenomenon on complex network.

\section{Conclusion}

In this paper, we propose a self-organized particle moving model
on scale free network with the shortest path algorithm and the
preferential walk algorithm. We study the number of over-capacity
vertices $A(t)$ varies as time $t$ and find that $A(t)$ forms the
punctuated equilibrium, that the over-capacity density
$\rho(A_{max})$ and $A_{max}$ have power law relationship which is
independent of the parameters n and a, these findings reflect the
interaction between the dynamical algorithm and the underlying
network structure. On the other hand, we analyze the temporal
correlation and the power spectrum of $A(t)$ and observe that our
model has the bell-shaped temporal correlation and the $1/f^{2}$
behavior, which may be caused by the preferential walk algorithm
and the underlying network structure but may have stronger
relationship with the topology structure of the underlying
network.

Finally, we study the average lifetime of particles accumulated
before time t. At the beginning, as the particles move along the
shortest path and the statistical property, the lifetime varies
slowly along with time t (see the part AB in Fig.5). As time
elapsing, more and more particles exist in the network and move
according to the preferential walk algorithm, which cause
particles move along the farther path thus lengthen the lifetime.

What's more, our work in this paper may guide the studying of
traffic phenomenon on complex networks, and we will further study
this subfield of dynamics on complex network in the future.

\end{document}